\documentclass[final,5p]{elsarticle}

\usepackage{acronym}
\usepackage{amsmath}
\usepackage{graphicx}
\usepackage[utf8]{inputenc}
\usepackage{mathtools}
\usepackage{multirow}
\usepackage{tabularx}
\usepackage{subcaption}

\newacro{ATL}{Anterior Temporal Lobe}
\newacro{CCA}{Canonical Correlation Analysis}
\newacro{CSC}{Controlled Semantic Cognition}
\newacro{EEG}{Electroencephalogram}
\newacro{EMC}{Embodied Music Cognition}
\newacro{fMRI}{functional Magnetic Resonance Imaging}
\newacro{GCCA}{Generalized Canonical Correlation Analysis}
\newacro{GloVe}{Global Vectors for Word Representation}
\newacro{LT243}{Language Topics (243 Sentences)}
\newacro{LT384}{Language Topics (384 Sentences)}
\newacro{MAP}{Mean Average Precision}
\newacro{MG}{Music Genres}
\newacro{NL}{Natural Language}
\newacro{NTTL}{Neural Theory of Thought and Language}
\newacro{PC}{Predictive Coding}
\newacro{SVM}{Support Vector Machine}

\journal{arXiv} 

\begin{document} 

\begin{frontmatter} 

\title{Low-dimensional Embodied Semantics for Music and Language}


\author[a,c]{Francisco Afonso Raposo} 
\author[a,c]{David Martins de Matos}
\author[b,c]{Ricardo Ribeiro}
\address[a]{Instituto Superior Técnico, Universidade de Lisboa, Av. Rovisco Pais, 1049-001 Lisboa, Portugal} 
\address[b]{Instituto Universitário de Lisboa (ISCTE-IUL), Av. das Forças Armadas, 1649-026 Lisboa, Portugal}
\address[c]{INESC-ID Lisboa, R. Alves Redol 9, 1000-029 Lisboa, Portugal}



\begin{abstract} 
Embodied cognition states that semantics is encoded in the brain as firing patterns of neural circuits, which are learned according to the statistical structure of human multimodal experience. However, each human brain is idiosyncratically biased, according to its subjective experience history, making this biological semantic machinery noisy with respect to the overall semantics inherent to media artifacts, such as music and language excerpts. We propose to represent shared semantics using low-dimensional vector embeddings by jointly modeling several brains from human subjects. We show these unsupervised efficient representations outperform the original high-dimensional \acs{fMRI} voxel spaces in proxy music genre and language topic classification tasks. We further show that joint modeling of several subjects increases the semantic richness of the learned latent vector spaces.
\end{abstract}


\end{frontmatter} 


\section{Introduction}

The current consensus in cognitive science defends that conceptual knowledge is encoded in the brain as firing patterns of neural populations \citep{kiefer2012}. Correlated patterns of experience trigger the firing of neural populations, creating neural circuits that wire them together. Neural circuits represent semantic inferences involving the encoded concepts and connect neurons responsible for encoding different modalities of experience, such as emotional, linguistic, visual, auditory, somatosensory, and motor \citep{kiefer2012,lakoff2014,pulvermuller2018,ralph2017}. Thus, cognition implies transmodal and distributed aspects that allow embodied meaning to be grounded, via semantic inferences which can be triggered by any concept, regardless of its level of abstraction. This multimodal semantics, which draws on supporting evidence from several disciplines, such as psychology and neuroscience \citep{desai2011,lakoff2012,thibodeau2013}, is appropriately termed ``embodied cognition''.

Since embodied cognition explains semantics to be biologically implemented by neural circuitry which captures the statistical structure of human experience, this motivates a statistical approach based on brain activity to computationally model semantics. That is, since patterns of neural activity reflect conceptual encoding of experience, then statistical models of neural activity (e.g., measured with \ac{fMRI}) will capture conceptual knowledge, i.e., semantics. While each individual brain is a rich source of semantic information, it is also biased according to the life experience of that individual, meaning that idiosyncratic semantic inferences are always triggered for any subject, which implies a ``noisy'' semantic system. We claim that richer shared semantic descriptions can be captured by jointly modeling the statistics of neural firing of several similarly stimulated brains.

In this work, we propose modeling shared semantics as low-dimensional encodings via statistical inter-subject brain agreement using \ac{GCCA}. We model both music and natural language semantics based on \ac{fMRI} recordings of subjects listening to music and being presented with linguistic concepts. By correlating the brain responses of several subjects to the same stimuli, we intend to uncover the latent shared semantics that are encoded in the \ac{fMRI} data representing human (brain) interpretation of media. Shared semantics refers to the shared statistical structure of the neural firing, i.e., the shared meaning of media artifacts, across a population of individuals. This is a useful concept for automatic inference of semantics from brain data that is meaningful for most people. We evaluate how semantic richness evolves with the number of subjects involved in the modeling process, via across-subject retrieval of \ac{fMRI} responses and proxy, downstream, semantic tasks for each modality: music genre and language topic classifications. Results show that these low-dimensional embeddings not only outperform the original \ac{fMRI} voxel space of several thousand dimensions, but also that their semantic richness improves as the number of modeled subjects increases.

The rest of this article is organized as follows: Section \ref{sec:related} reviews related work on embodied cognition, both in terms of music and natural language, as well as computational approaches leveraging some of its aspects; Section \ref{sec:setup} describes our experimental setup; Section \ref{sec:results} presents the results; Section \ref{sec:discussion} discusses the results; and Section \ref{sec:conclusions} draws conclusions and considers future work.

\section{Related work}
\label{sec:related}

Embodied cognition has been accumulating increasing amounts of supporting evidence in cognitive science \citep{kiefer2012,lakoff2012}. This research direction was initially motivated by the fact that embodied cognition is able to account for the semantic grounding of natural language. Grounding is rooted in embodied primitives, such as somatosensory and sensorimotor concepts \citep{lakoff2014}, which implies humans ultimately understand high-level (linguistic) concepts in terms of what their mediating physical bodies afford in their physical environment. The \ac{NTTL} is an embodied cognition framework proposed by \citet{lakoff2012}, which casts semantic inference as ``conceptual metaphor''. This is based on the observation that metaphorical thought and understanding are independent of language. Linguistic and other kinds of metaphors (e.g., gestural and visual) are seen as surface realizations of conceptual metaphors, which are realized in the brain via neural circuits learned according to repeated patterns of experience. For instance, the abstract concept of ``communicating ideas via language'' is understood via a conceptual metaphor: ideas are objects; language is a container for idea-objects; and communication is sending idea-objects in language-containers. This metaphor maps a source domain of sending objects in containers to a target domain of communicating ideas via language \citep{lakoff2014}. \citet{ralph2017} proposed the \ac{CSC} framework which, much like \ac{NTTL}, features multimodal and distributed aspects of cognition, i.e., semantics is also defined by transmodal circuits connecting different modality-specific neuron populations distributed across the whole brain. It contrasts with \ac{NTTL}, however, by proposing the existence of a transmodal hub in the \acp{ATL}. \citet{pulvermuller2018} introduced the concepts of semantic ``kernel'', ``halo'', and ``periphery'' in order to address different levels of generality of conceptual semantic features. The periphery contains information about very idiosyncratic features and referents, whereas the kernel contains the most generic features of a concept. The generality level of the halo features fall in between the levels of the kernel and periphery. These correspond to larger (periphery) to smaller (kernel) sets of neurons encoding semantic information about concepts and are merely approximations of what is likely to be a smooth continuum of increasingly more generic (and smaller set size of) neurons. Hebbian (and anti-Hebbian) learning is what allows ``semantic feature extraction'' (conceptual metaphor) to happen. Conceptual flexibility is proposed to be implemented via neurobiological mechanisms such as priming and gain control modulation \citep{pulvermuller2018}. A review of several families of conceptual knowledge encoding theories is presented by \citet{kiefer2012}, where the authors conclude by arguing in favor of the embodiment family.

Even though embodied cognition was initially motivated by the acquisition of a deeper understanding of natural language semantics, it is inherently multimodal and generic. Therefore, it also accounts for music semantics and, indeed, some work has already been published following this cognitive perspective. \citet{kreyn2018} emphasizes the role of modulations of physical tension in music semantics. These modulations are conveyed by tonal and temporal relationships in music artifacts, consisting of manipulations of tonal tension in a tonal framework (musical scale). This embodied perspective is reinforced by the fact that tonal perception seems to be biologically driven, since one-day old babies physically perceive tonal tension \citep{virtala2013}. This is thought to be a consequence of the ``principle of least effort'', where consonant sounds, which consist of more harmonic overtones, are more easily, i.e., efficiently, processed and compressed by the brain than dissonant sounds, creating a more pleasant experience \citep{bidelman2009}. \citet{leman2010} also claims music semantics is defined by the mediation process of the listener, i.e., the human body and brain are responsible for mapping from the physical modality (audio) to the experienced modality. His theory, \ac{EMC}, also supports the idea that semantics is motivated by affordances, i.e., music is understood in a way that is relevant for functioning in a physical environment. \citet{koelsch2019} proposed the \ac{PC} framework, also pointing to the involvement of transmodal neural circuits in both prediction and prediction error resolution of musical content. This process of active inference makes use of ``mental action'' while listening to music, thus, also pointing to the role of both action and perception in semantics.

Given the growing consensus that brain activity is a manifestation of semantic inferences, it is only natural to model it computationally. In particular, statistical modeling seems appropriate, given the seemingly stochastic nature of human brain dynamics. The most straightforward way to capture embodied semantics, which implies that conceptual knowledge is spatially encoded across the brain, is to model \ac{fMRI} data, which consist of volume readings of brain activity, thus, providing a direct reading from a specific spatial location. This makes the decoding of such data easier than, for instance, decoding \acp{EEG}, which are scalp measurements of electrical activity, i.e., the readings of each scalp sampling location are result from aggregated activity originating from many distributed spatial locations in the brain. Accordingly, \citet{pereira2018} proposed a method for decoding linguistic meaning from \ac{fMRI}, using ridge regression \citep{hoerl1970} to learn a mapping from \ac{fMRI} to \ac{GloVe} embeddings \citep{pennington2014}, which are statistically learned to describe textual meaning. Within-subject models were significantly able to generalize to new concepts. More importantly, those models were able to generalize to whole sentences, even though they were only trained with single concepts. The authors found that the most informative voxels were widely distributed across the brain, as predicted by embodied cognition. \citet{casey2017} performed analogous experiments with \ac{fMRI} responses to short music clips spanning five music genres. Stimuli and genre \ac{SVM} \citep{cortes1995} classification results significantly outperformed the random baselines. Furthermore, melodic features were extracted from the stimuli and used for ridge regression of voxel sphere responses, also achieving significant results for some brain regions.

Our approach differs from previous computational approaches in the following ways: we never use vector space descriptions extracted from stimuli data in our models, as opposed to, for instance, \ac{fMRI}-based \ac{GloVe} regression \citep{pereira2018} or melody-based voxel sphere regression \citep{casey2017}; and we do not constrain our semantic vector space learning to any taxonomy, as opposed to, for instance, \ac{fMRI}-based genre classification \citep{casey2017}. These differences allow for learning a generic semantic vector space, which can later be tested by semantic proxy tasks to assess its semantic richness.

\section{Experimental setup}
\label{sec:setup}

In this section, we describe the data that was used in our experiments (Section \ref{sub:datasets}), the features and preprocessing (Section \ref{sub:features}), the \ac{GCCA} model (Section \ref{sub:gcca}), and the evaluation setups: across-subject retrieval (Section \ref{subsub:retrieval}), and classification (Section \ref{subsub:classification}).

\subsection{Datasets}
\label{sub:datasets}

We experiment with three datasets, taken from two published datasets \citep{hanke2015,pereira2018}. \citet{hanke2015} had 20 human subjects listen to 25 short music clips (6s) spanning 5 music genres: ambient, country, metal, rocknroll, and symphonic. We select the data for every subject that listened to each clip 8 times (all but one), yielding a final dataset of 19 subjects and 75 clip segments (\ac{fMRI} data is sampled every 2s seconds), hereafter, referred to as \ac{MG}. \citet{pereira2018} presented 180 linguistic concepts to 16 subjects using 3 different presentation paradigms, 8 of whom were presented with additional 243 sentences spanning 24 topics, and 6 of whom were presented with additional 384 sentences spanning another 24 topics. We select the two sentence subsets for evaluation (even though we also use the average concepts data for training the embedders) since they provide topic labels for proxy task evaluation and, hereafter, we refer to them as \ac{LT243} and \ac{LT384}. Table \ref{tab:topics} lists the topics.

\begin{table}[htb]
\small
\begin{center}
\caption{Language topics}
\label{tab:topics}
\begin{tabular}{c|c}
\ac{LT243} & \ac{LT384}\\
\hline
Astronaut & Animal\\
Beekeeping & Appliance\\
Blindness & Bird\\
Bone Fracture & Body Part\\
Castle & Building Part\\
Computer Graphics & Clothing\\
Dreams & Crime\\
Gambling & Disaster\\
Hurricane & Drink Non Alcoholic\\
Ice Cream & Dwelling\\
Infection & Fish\\
Law School & Fruit\\
Lawn Mower & Furniture\\
Opera & Human\\
Owl & Insect\\
Painter & Kitchen Utensil\\
Pharmacist & Landscape\\
Polar Bear & Music\\
Pyramid & Place\\
Rock Climbing & Profession\\
Skiing & Tool\\
Stress & Vegetable\\
Taste & Vehicles Transport\\
Tuxedo & Weapon\\
\end{tabular}
\end{center}
\normalsize
\end{table}

We randomly partition each dataset into folds based on stimuli, for use in cross-validation evaluation setups, in a way that guarantees at least one instance of each class in the test set and such that all folds have the same sizes. It is of the utmost importance to guarantee that train/test splits of recorded brain data are not biased in terms of local temporal coherence: preliminary experiments revealed that recorded brain activity exhibits a great degree of local temporal similarity that is not the result of similar stimuli processing and integration but that it rather reflects an overall short term temporally coherent brain state. For instance, if a sequence of sentences is presented to a subject (even if there are short time intervals between them), the \ac{fMRI} data corresponding to sentences that were shown consecutively will be at least among the most similar pairs of \ac{fMRI} vectors regardless of the content of the sentences. Consequently, in order to evaluate \ac{fMRI}-based classification, consecutive same-class stimuli instances cannot be in both the train and test splits, so as to guarantee the model learns to extract stimuli-related semantic content instead of general brain state content. Since music clips presentation in \ac{MG} is appropriately randomized, we need only to guarantee that all segments (three) from each clip are either in the train or test set. For \ac{LT243} and \ac{LT384}, the presentation of passages was also randomized but its sentences were not, meaning that we allocate every sentence from each passage either to the train or test set. Given the aforementioned constraints, \ac{MG}, \ac{LT243}, and \ac{LT384} are partitioned into five, three, and four folds, respectively.

\subsection{Features and preprocessing}
\label{sub:features}

Even though the original \ac{fMRI} datasets were recorded using different equipment (e.g., recordings by \citet{hanke2015} have a greater spatial resolution than the recordings by \citet{pereira2018}), they all roughly cover the whole brain, i.e., there is no informed brain area voxel selection. As embodied cognition posits that the semantic network spans the whole brain, this is interesting as it allows us to evaluate whether \ac{GCCA} is able to filter relevant activity in an end-to-end fashion. Since \ac{MG} has very high spatial resolution, the number of voxels is prohibitively large (both in terms of memory usage and processing time). Therefore, we downsample \ac{MG}, using the Nilearn library \citep{abraham2014}, by a factor of 6, resulting in 5488 voxels. Furthermore, we also average all runs in \ac{MG} in order to produce a single \ac{fMRI} per subject and stimulus.

\subsection{\acf{GCCA}}
\label{sub:gcca}

\ac{CCA} \citep{hotelling1936} is a method that finds $C\leq\operatorname{min}\left(d_0,d_1\right)$ pairs of maximally correlated linear projections $p_0^c$ and $p_1^c$ of two input views $X_0\in{\rm I\!R}^{d_0}$ and $X_1\in{\rm I\!R}^{d_1}$, such that the canonical dimensions are uncorrelated with each other. Formally:
\begin{align}
\left(p_0^*,p_1^*\right)
&=\underset{p_0\in{\rm I\!R}^{d_0},p_1\in{\rm I\!R}^{d_1}}{\operatorname{argmax}}\operatorname{corr}\left(p_0^{\bf{T}}X_0,p_1^{\bf{T}}X_1\right)\\
&=\underset{p_0\in{\rm I\!R}^{d_0},p_1\in{\rm I\!R}^{d_1}}{\operatorname{argmax}}\frac{p_0^{\bf{T}}\Sigma_{01}p_1}{\sqrt{p_0^{\bf{T}}\Sigma_{00}p_0p_1^{\bf{T}}\Sigma_{11}p_1}}
\end{align}
where $\Sigma_{00}$ and $\Sigma_{11}$ are the covariances of $X_0$ and $X_1$, respectively, and $\Sigma_{01}$ is the cross-covariance. \ac{GCCA} \citep{horst1961,kettenring1971} is an extension of \ac{CCA} for an arbitrary number of views $V$. Formally, it minimizes:
\begin{equation}
\sum_{v=1}^{V}{||G-X_vP_v||_F^2}
\end{equation}
where $G^{\bf{T}}G=I_C$, $P_v\in{\rm I\!R}^{d_v\times{C}}$ is the projection matrix for view $v$ (stacking $C$ linear projections), $X_v\in{\rm I\!R}^{N\times{d_v}}$ is the data matrix for view $v$, and $N$ is the number of samples. $G$ is the canonical vector space shared by all $V$ views. This is the space that captures the latent semantics that are uncovered via correlating brain (\ac{fMRI}) views. \ac{CCA}-based models have previously been used for learning cross-modal semantic spaces which essentially leveraged the transmodal aspect of human semantic cognition as manifested in multimodal artifacts. For instance, \citet{raposo2019} modeled correlations between music audio and dance video and \citet{yu2019} modeled correlations between music audio and lyrics text.

\subsection{Evaluation}
\label{sub:evaluation}

We start by determining the appropriate number of semantic dimensions (canonical components) for each dataset. For each tested number of components (from 2 to 25), we train the \ac{GCCA} models in a cross-validation setup and evaluate its generalizability by computing the \ac{MAP} on the test set in an across-subject retrieval of \ac{fMRI} task. Then, we select the number of components that maximizes this metric (averaged over all folds). After deciding on the appropriate number of semantic dimensions, we evaluate the semantic richness of this latent space via appropriate proxy tasks: music genre classification and language topic classification. These tasks are evaluated for a variable number of modeled subjects by \ac{GCCA}, in order to study how semantic richness evolves with an increasing number of subjective brain responses. Since there is more than one way of choosing a variable number of subjects to model out of all subjects, we randomly sample up to 50 combinations, run the experiments for each combination, and report the average results.

\subsubsection{Across-subject retrieval}
\label{subsub:retrieval}

This task consists of retrieving a sorted list of \ac{fMRI} database items, given an \ac{fMRI} query, based on the cosine similarity between semantic vectors. The database contains all test set \ac{fMRI} responses from all subjects besides the query itself. We compute \ac{MAP}, which is appropriate to assess ranking quality in sorted lists of database items which contain more than a single relevant item. The relevance criterion considers any \ac{fMRI} response from other subjects, for the same stimulus as the \ac{fMRI} query, to be relevant. We also compute random performance baselines.

\subsubsection{Music genre and language topic classifications}
\label{subsub:classification}

Both music genre and language topic classification tasks consist of classifying query items into one of the classes modeled by a classifier. We use \acp{SVM} for classifying music clips into 1 of 5 music genres and language sentences into 1 of 24 topics for both language datasets. The data instances used for training and testing the \acp{SVM} are the average (over the corresponding modeled subjects) of the \ac{GCCA} projections of the corresponding \acp{fMRI}, which represent the semantic embeddings as computed by the \ac{GCCA} model. We also compute two baselines: the average classification performance over all subjects using the original high-dimensional (i.e., several thousands of voxels) \ac{fMRI} data; and random performance. We perform classification in a grid search setup for parameters $C$ and $\gamma$ and report the results for a particular combination that maximizes the average classification accuracy across folds.

\section{Results}
\label{sec:results}

The optimal number of dimensions (which are chosen for the proxy classification tasks) are 9, 11, and 6 for \ac{MG}, \ac{LT243}, and \ac{LT384}, respectively, as determined via the across-subject retrieval task, which yielded \ac{MAP} scores of 0.272, 0.070, and 0.044 for \ac{MG}, \ac{LT243}, and \ac{LT384}, respectively (all statistically significant results according to randomization tests against the random baseline performances \citep{bestgen2015} of 0.081, 0.022, and 0.017, respectively).

\begin{figure}[!h]
\begin{subfigure}{\linewidth}
\includegraphics[width=\linewidth]{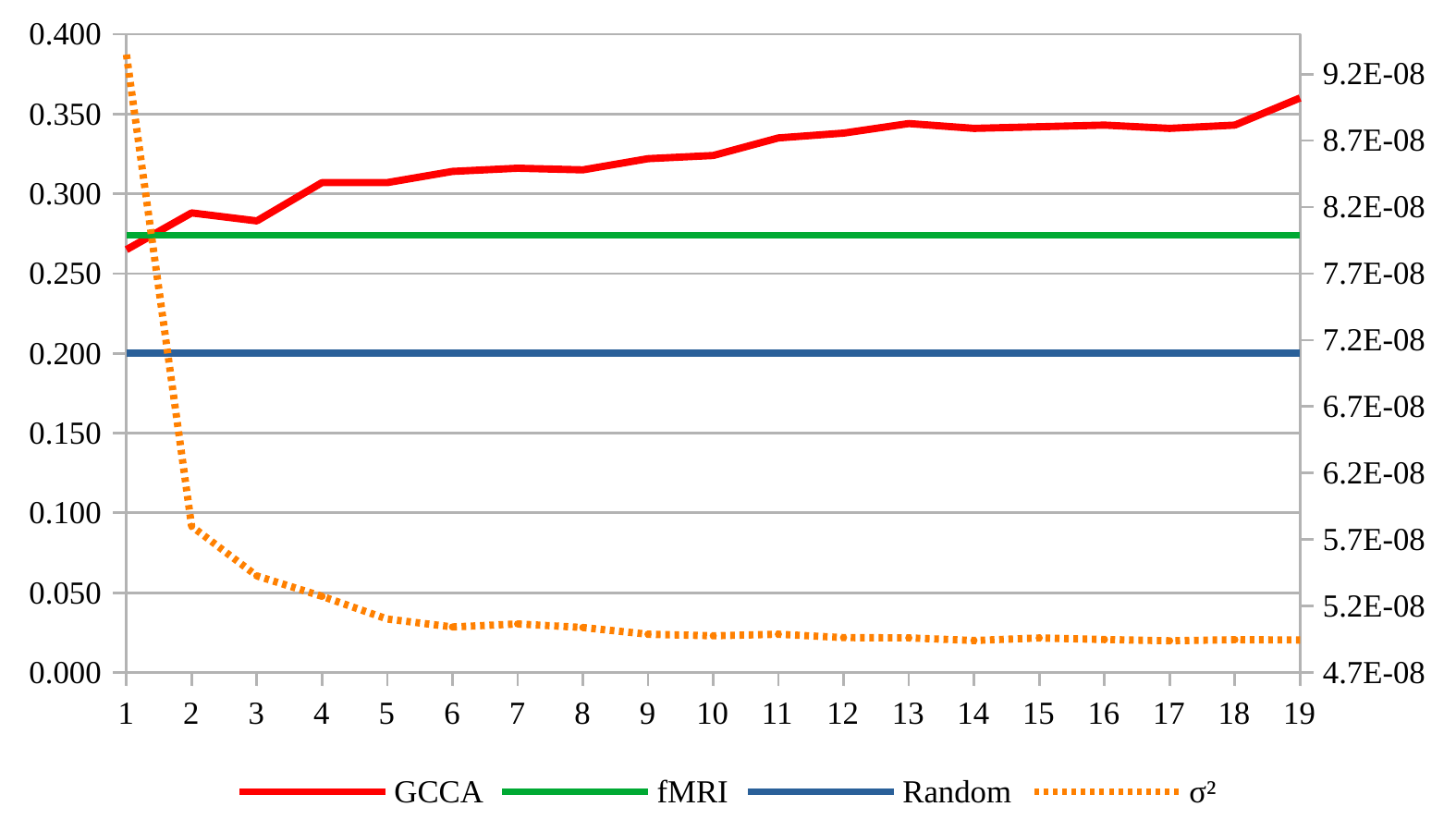}
\caption{\ac{MG}}
\label{fig:mg}
\end{subfigure}
\begin{subfigure}{\linewidth}
\includegraphics[width=\linewidth]{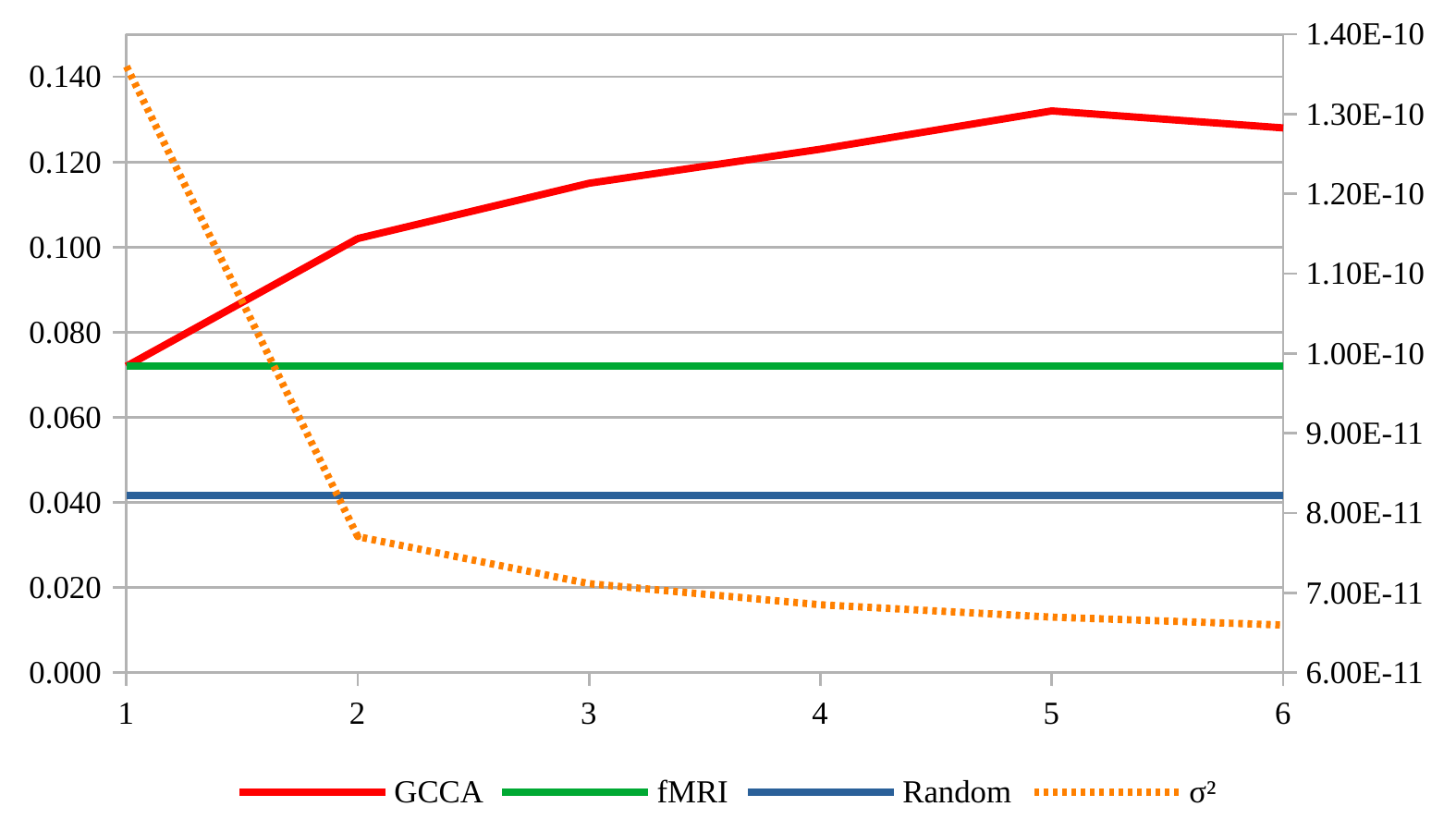}
\caption{\ac{LT243}}
\label{fig:lt243}
\end{subfigure}
\begin{subfigure}{\linewidth}
\includegraphics[width=\linewidth]{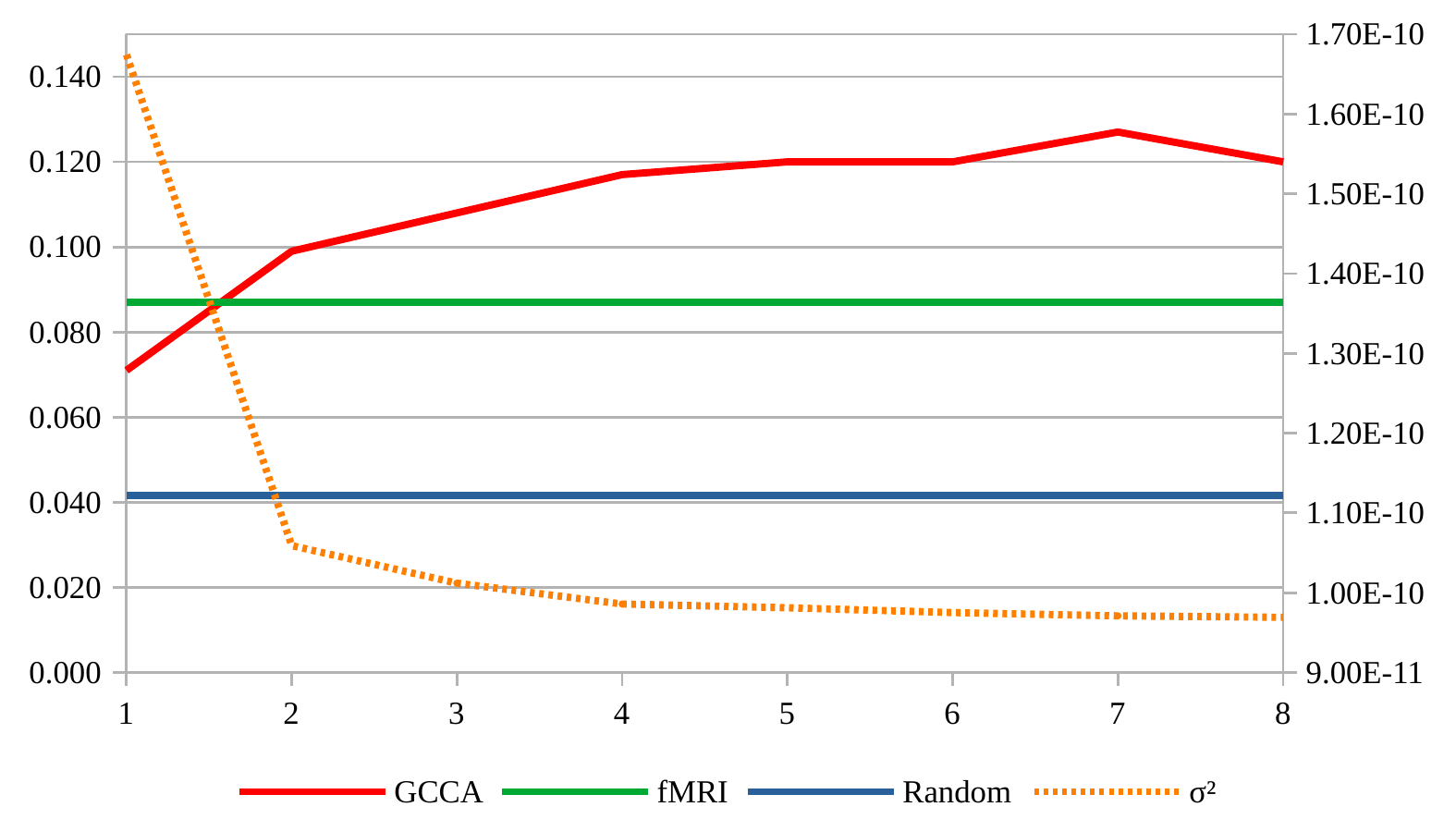}
\caption{\ac{LT384}}
\label{fig:lt384}
\end{subfigure}
\caption{Classification accuracies (red, green, and blue lines relative to left y-axes) and voxel score distribution variances (orange dashed lines relative to right y-axes) vs. number of modeled subjects.}
\label{fig:classification}
\end{figure}

Figure \ref{fig:mg} shows the evolution of music genre classification performance for an increasing number of modeled subjects in \ac{MG} (red, green, and blue lines relative to the left y-axis). The best accuracy score is 0.360. Figures \ref{fig:lt243} and \ref{fig:lt384} show the evolution of language topic classification performance for an increasing number of modeled subjects in \ac{LT243} and \ac{LT384} (red, green, and blue lines relative to the left y-axes), respectively. The best accuracy scores are 0.132 and 0.127 for \ac{LT243} and \ac{LT384}, respectively.




\section{Discussion}
\label{sec:discussion}

We assessed statistical significance on the across-subject retrieval task \ac{MAP} scores via randomization tests for 32768 randomizations and obtained p-values $p<7\times 10^{-5}$, $p<3\times 10^{-5}$, and $p<\times 10^{-5}$ for \ac{MG}, \ac{LT243}, and \ac{LT384}, respectively. We also ran binomial statistical significance tests for the classification performances. \ac{fMRI}-based classification in \ac{MG} did not significantly differ from random performance. \ac{GCCA} embeddings are significantly better than random performance starting at 4 subjects ($1\times 10^{-3}<p<3\times 10^{-2}$). \ac{fMRI}-based classification in \ac{LT243} and \ac{LT384} is significantly better than random performance, with p-values $p<4\times 10^{-2}$ and $p<2\times 10^{-4}$, respectively. \ac{GCCA}-based classification in \ac{LT243} and \ac{LT384} is significantly better than random performance for any number of subjects and is also significantly better than \ac{fMRI}-based classification starting from 3 ($1\times 10^{-3}<p<3\times 10^{-2}$) and 5 ($1\times 10^{-2}<p<3\times 10^{-2}$) subjects, respectively. 

The fact that the retrieval tasks showed generalizability of the model is already evidence that it captures some semantic aspects of the stimuli, since the only supervision during training is (musical and linguistic) stimulus-based matching between \ac{fMRI} data from different subjects. Proxy tasks results further validate this claim by showing the predictive power of the shared semantics in terms of specific semantic concepts (music genres and language topics). These tasks are appropriate for evaluating shared semantics because they involve inference of semantic concepts which are stable across individuals. Recall that the learned embeddings are generic, i.e., not specifically optimized for these concepts, yet they performed remarkably well in these tasks, meaning that these shared concepts were uncovered by jointly modeling inter-subject \ac{fMRI} data in an unsupervised fashion. Moreover, these low-dimensional embeddings significantly outperformed the \ac{fMRI} spaces using hundreds and thousands of times less dimensions, showing their efficiency.

These semantics were shown to be improved by including additional subjects to the model, suggesting that the semantics of the stimuli is indeed latent in the \ac{fMRI} data, but also that it is polluted by noisy inferences which can be better filtered out by \ac{GCCA} when more views are taken into account. It is likely that modeling less subjective semantic systems, i.e., brains, is not enough to filter out the conceptual semantic peripheries (using the terminology of \citet{pulvermuller2018}) and other unrelated but co-occurring (during \ac{fMRI} data acquisition) brain activity, which are both non-robust semantic descriptions of the concepts encoded in the media artifacts used as stimuli. It is equally possible that modeling more brains helps the model in finding semantically robust voxels, i.e., voxels belonging to the semantic kernel and halo, that would not be found otherwise. In order to get some insight on which of these effects is actually happening in the context of these experiments, we ran an additional analysis that measures how the \ac{GCCA} scores are distributed across voxels, regardless of where those voxels are located. For each number of modeled subjects, we compute the distribution of voxel scores, normalize them in order to sum to 1, and compute the average variance of these distributions across folds and subject combinations. A higher value means that the voxel scores are concentrated in a smaller number of voxels, whereas a lower value means that the voxels scores are concentrated in a larger number of voxels. Voxel scores are computed according to $\operatorname{score}_v\left(i\right)=\sqrt{\sum_{c=0}^{C-1}P_v^2\left(i,c\right)}$, where $v$ is the subject (view) index, $i$ is the voxel index and $c$ is the canonical dimension index. The orange dashed lines (relative to the right y-axes) in Figures \ref{fig:mg}, \ref{fig:lt243}, and \ref{fig:lt384} illustrate the evolution of voxel score distribution variance for \ac{MG}, \ac{LT243}, and \ac{LT384}, respectively.

The voxel score distribution variance plots suggest that the model is able to find more semantically rich voxels as the number of modeled subjects increases. Note, however, that these experiments were performed with a relatively small number of samples (a few hundreds) compared to the number of input dimensions (several thousands) being modeled. This means that there is an upper bound on the number of latent dimensions that \ac{GCCA} can model for each view. Therefore, adding additional subjective views, in these conditions, seems to help in finding the semantic kernel and halo, as if these views are artificially boosting the sample size (number of stimuli). It is still possible that the inverse effect (filtering out of the semantic periphery) can happen with a larger sample size and that is a direction for further investigation. Finally, it is interesting that all these effects are observed for both music and language modalities. These results not only show the promise of \ac{CCA}-based methods to model semantics via brain dynamics but also suggest that meaning in music and language is inferred in a similar way, thus, providing further evidence for embodied cognition in general.

\section{Conclusions and future work}
\label{sec:conclusions}

We showed that music and language semantics can be learned and studied via canonically correlating \ac{fMRI} responses from different subjects. \ac{GCCA} models were able to generalize how the brains from all subjects covary with respect to musical and linguistic stimuli as well as to produce unsupervised semantic embeddings for classification of music genres and language topics. Moreover, we showed that the semantic space learned by \ac{GCCA} is more powerful as the number of modeled subjects increases. Future work includes experimenting with larger and more varied datasets, using \ac{GCCA} to study which brain areas covary together for both music and language domains, as well as extending this study to other modalities.

\bibliographystyle{elsarticle-num-names} 
\bibliography{lowdimsem}

\end{document}